# Towards Ultimate Memory with Single-Molecule Multiferroics

Yali Yang[1,2,3,†], Liangliang Hong[2,3,†], Laurent Bellaiche[4], and Hongjun Xiang[2,3,*]

[1]School of Mathematics and Physics, University of Science and Technology Beijing, Beijing 100083, China

[2]Key Laboratory of Computational Physical Sciences (Ministry of Education), State Key Laboratory of Surface Physics, and Department of Physics, Fudan University, Shanghai 200433, China

[3]Shanghai Qi Zhi Institute, Shanghai 200030, China

[4]Physics Department and Institute for Nanoscience and Engineering, University of Arkansas, Fayetteville, Arkansas 72701, USA

**Abstract**

The demand for high-density storage is urgent in the current era of data explosion. Recently, several single-molecule (-atom) magnets/ferroelectrics have been reported to be promising candidates for high-density storage. As another promising candidate, single-molecule multiferroics are not only small but also possess ferroelectric and magnetic orderings, which can sometimes be strongly coupled and used as data storages to realize the combination of electric writing and magnetic reading. However, they have been rarely proposed, and never been experimentally reported. Here, by building Hamiltonian models, we propose a new model of single-molecule multiferroic in which electric dipoles and magnetic moments are parallel and can rotate with the rotation of the single molecule. Furthermore, with performing spin-lattice dynamics simulations, we reveal the conditions (*e.g.*, large enough single-ion anisotropy and appropriate electric field) under which the new single-molecule multiferroic can arise. Based on this model, as well as first-principles calculations, a realistic example $Co(NH_3)_4N$@SWCNT is constructed and numerically confirmed to demonstrate the feasibility of the new single-molecule multiferroic model. Our work not only sheds light on the discovery of single-molecule multiferroics but also provides a new guideline to design multifunctional materials for ultimate memory devices.

**Introduction**

With the relentless pursuit of high-performance electronic devices and the rapid advancement of technology, the augmentation of data storage capacity has emerged as a pivotal objective in the development of electronic devices. Substantial strides have been achieved in this domain over the past few decades[1-4]. For example, considering the fact that the ferroelectric (FE) and magnetic orderings can coexist or even be strongly coupled in multiferroics, multiferroic random access memories (MFRAMs) have been designed and fabricated to not only achieve non-volatility but also offer the possibility of combining the advantages of ferroelectric and magnetic access memories, *i.e.*, electric writing combined with magnetic reading[5-9]. In particular, during writing, applying an electric field is generally easier and less energy-costly than applying a magnetic field, while magnetic reading is faster, non-destructive, and has no wear-out mechanism as compared to electric reading in storage[1].

The demand for high-density storage requires the storage units to be increasingly smaller. However, in conventional storage, the size of the storage material is relatively large, making it difficult to meet the increasing demands of high-electronic device integration. Therefore, in both the design of the device structure and the quest for optimal storage materials, the smaller the storage unit, the better. As good candidates, single atoms or molecules could lead to the shrinking of nonvolatile memory devices[10-15]. The usage of single-atom (-molecule) magnets, or single-molecule ferroelectrics for designing high-density storage has made some progress. For example, a very interesting research work was recently reported by Zhang *et al.* where they observed a gate-controlled switching between two electronic states in single-molecule electret Gd@C82[16]. More specifically, the encapsulated Gd atom serves as a charged center, establishing two distinct single-electron transport channels. This configuration enables the system to exhibit a ferroelectricity-like hysteresis loop, rendering it highly promising for future miniaturized storage devices. Compared with single-atom (-molecule) ferroelectrics and magnets, single-molecule multiferroics are also small in size, however, they can achieve the coexistence and even coupling of electric dipoles and magnetism. Consequently, if they can be used as a single memory unit, one can not only move towards ultimate memory but also achieve electric writing and magnetic reading, which would lead to higher data storage efficiency. For example, recently, Zeng *et al.* theoretically designed a single-molecule multiferroic by intercalating the buckled metal porphyrin molecules in a bilayer of 2D materials. The 3d metal ion in the metal molecule can have strong ionic/covalent binding interaction with the surface anions of the 2D materials and, under such circumstances, the magnetization distribution or direction of metal porphyrin can be switched upon FE switching[15]. This mechanism relies on the strong interaction between the 2D materials and magnetic single-molecule. To the best of our knowledge, single-molecule multiferroics have rarely been proposed theoretically, and have not been yet reported experimentally.

In this work, we first propose a new simplified single-molecule multiferroic model and perform spin-lattice dynamics simulations based on a simple Hamiltonian model. We reveal that single-molecule multiferroicity can arise in systems for which the molecules have intrinsic electric dipoles and large single-ion anisotropy. The magnetic moment of the molecule reverses simultaneously as the molecule rotates under electric fields. Then, a realistic system is constructed, in combination with first-principles calculations, to verify the new single-molecule multiferroic model we presently propose.

**Methods**

All density functional theory (DFT) calculations are performed with the Vienna *ab*-initio Simulation Package (VASP)[17]. The generalized gradient approximation (GGA)[18] in the form of the Perdew-Burke-Ernzerhof (PBE) functional and the projector augmented wave (PAW) method[19] are applied to describe the exchange-correlation and the core electrons, respectively. The GGA+U method[20] with $U_{eff}$ = 3 eV is employed to treat the d orbitals of transition metals. The plane-wave cutoff and convergence criteria for energy and force are set to be 520 eV, $10^{-6}$ eV, and 0.01 eV/Å, respectively. The spin-dynamics simulations are performed with the package PASP (Property Analysis and Simulation Package for materials)[21].

**Results and discussion**

*General idea.* –To achieve the goal of single-molecule multiferroics and ultimate memory, a new scheme is proposed here with the multiferroic molecules acting as the smallest storage units and be orderly distributed throughout the system. As shown in Figure 1, we use the purple ovals to indicate the multiferroic molecules, inside which the blue and red arrows represent the electric dipoles and magnetic moments, respectively. Here, the electric dipole of each molecule should be parallel to its magnetic easy axis, otherwise, the electric manipulation of the magnetization cannot be achieved. Furthermore, we assume that the single ion anisotropy of the molecule is large enough so that the magnetic moment of the molecule can rotate simultaneously with the electric dipole. Taking advantage of the unique property of the multiferroic molecule, we can achieve a switch between different states through electrically-controlled magnetism without applying a magnetic field, thereby realizing high-data-density and low-energy-consuming storage.

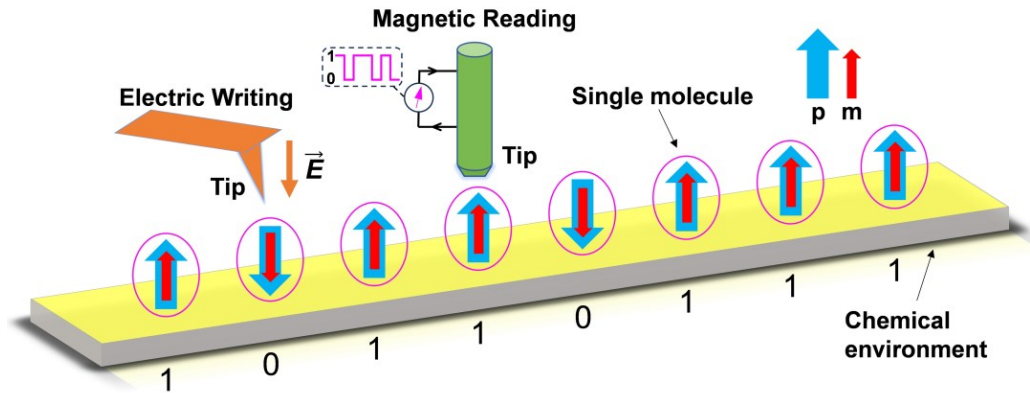

Figure 1. Schematic of the single-molecule multiferroics device proposed in this work. The purple ovals with blue and red arrows inside represent the multiferroic single molecules. The blue and red arrows indicate the electric dipole ($p$) and magnetic moment ($m$), respectively.

In order to clarify the new scheme of single-molecule multiferroics more intuitively, we design a simple model shown in Figure 2a. The model includes a diatomic molecule which is indicated by the yellow and blue balls and two neutral atoms which are indicated by the black balls, representing the chemical environment. We note that the diatomic molecule is able to rotate but with its barycenter pinned at the position marked by the red cross in Figure 2a. In order to make the model simpler, we restrict the rotation of the diatomic molecule to two dimensions, making the rotation angle $\theta$ the only generalized coordinate of the molecule. More importantly, the diatomic molecule is designed to have both an electric dipole and magnetic moment (*i.e.*, $p$ and $m$, respectively), which makes it possible for the appearance of multiferroicity. The angle between the electric dipole and magnetic moment is labeled by $\varphi$. In our model, an external electric field is applied to the system along the $z$-direction.

To further investigate the possibility and decisive factors of the system becoming a multiferroic, a Hamiltonian model is constructed and then spin-lattice dynamic simulations are performed for the system. In general, the Hamiltonian consists of the following four parts

$$H = E_{vdW} + E_{dipole} + E_{SIA} + E_{rot}, \qquad (1)$$

The first term $E_{vdW}$ represents the van der Waals (vdW) interaction between the neutral atoms and the diatomic molecule. Here, we adopt the attraction part of the Lennard-Jones potential to approximate this term,

$$E_{vdW} = -A\left(\frac{1}{t_1^6} + \frac{1}{t_2^6} + \frac{1}{s_1^6} + \frac{1}{s_2^6}\right), \ (A>0) \qquad (2)$$

where $t_1$, $t_2$, $s_1$, and $s_2$ are the distance between atoms indicated in Fig. 2a. $A$ is the potential parameter and is defined to be positive here. It is easy to see that besides the distance between the diatomic molecule and the surrounding atoms, the magnitude of $A$ also determines the vdW attraction energy of the system, which indicates that a suitable ferroelectric reversal barrier could be achieved if the system presents a suitable magnitude of $A$. The second term $E_{dipole}$ describes the interaction between the external electric field and the electric dipole moment, which has the explicit form:

$$E_{dipole} = -p \cdot E \cos\theta. \qquad (3)$$

The third term $E_{SIA}$ originates from the single-ion anisotropy of the magnetic ion in the diatomic molecule, and it can be expressed as

$$E_{SIA} = -D(m_x \sin\theta + m_z \cos\theta)^2, \qquad (4)$$

where $m_x$ and $m_z$ are the components of the magnetic moment, and $D$ is the parameter of the single-ion anisotropy. The last term $E_{rot}$ involves the rotation of the diatomic molecule, and can be written as

$$E_{rot} = \frac{1}{2}I\omega^2, \qquad (5)$$

where $I = m_1 l_1^2 + m_2 l_2^2$ and $\omega = \frac{d\theta}{dt}$. The parameters involved in the Hamiltonian are shown in Figure 2a.

After the introduction of the Hamiltonian model, we come to the stage of spin-lattice dynamics simulations. The modified Euler's method and the semi-implicit method devised by Mentink et al.[22] are adopted to numerically solve the Newton's equation and the Landau–Lifshitz–Gilbert equation,

$$(m_1 l_1^2 + m_2 l_2^2)\frac{d^2\theta}{dt^2} = M - C\frac{\partial \theta}{\partial t}, \quad M = -\frac{\partial H}{\partial \theta}, \quad (6)$$

$$\frac{d\vec{m}}{dt} = -\gamma_L \vec{m} \times \vec{B}_{eff} - \gamma_L \frac{\alpha}{|\vec{m}|}\vec{m} \times (\vec{m} \times \vec{B}_{eff}), \quad \vec{B}_{eff} = -\frac{\partial H}{\partial \vec{m}}, \quad (7)$$

where $C$ is the damping coefficient and $B_{eff}$ is the effective magnetic field. We note that the spin-lattice dynamics method is now integrated into the software package PASP (Property Analysis and Simulation Package for materials)[21]. Part of the parameters used in the simulations are shown in Table I. We note that some parameters are assumed without considering their physical realities, such as the magnetic moment $m$ and the electric dipole $p$. As to the parameters, e.g., $D$ and $E$, which are not listed in Table I, we adjust their values during the simulations to investigate how the model behaves under different conditions. In our simulations, the initial value of $\theta$ is set to be 15°, which can be viewed as a result of thermodynamic fluctuations in real situations. Besides, the initial direction of the magnetic moment is set to be aligned with the electric dipole moment. It is also worth noting that Eqs. (6) and (7) can only simulate the system at zero temperature. The simulations at finite temperatures are performed by modifying the dynamical equations with Langevin thermostats [see details in Section IV of the Supplementary Materials (SM)].

TABLE I. The parameters of the simple model adopted in the spin-lattice dynamic simulations. We note that $\theta_0$ denotes $\theta_{t=0}$ and $\delta t$ represents the time step of the dynamic simulations.

| Parameter | $l_1$ | $l_2$ | $d_1$ | $d_2$ | $m_1$ | $m_2$ | $A$ |
|---|---|---|---|---|---|---|---|
| Value | 1.0 | 1.0 | 3.0 | 3.0 | 1.0 | 1.0 | 10.0 |
| Parameter | $p$ | $M$ | $\gamma_L$ | $\alpha$ | $\theta_0$ | $\delta t$ | $C$ |
| Value | 1.0 | 1.0 | 1.0 | 0.1 | 15.0 | 5.0E-4 | 0.8 |

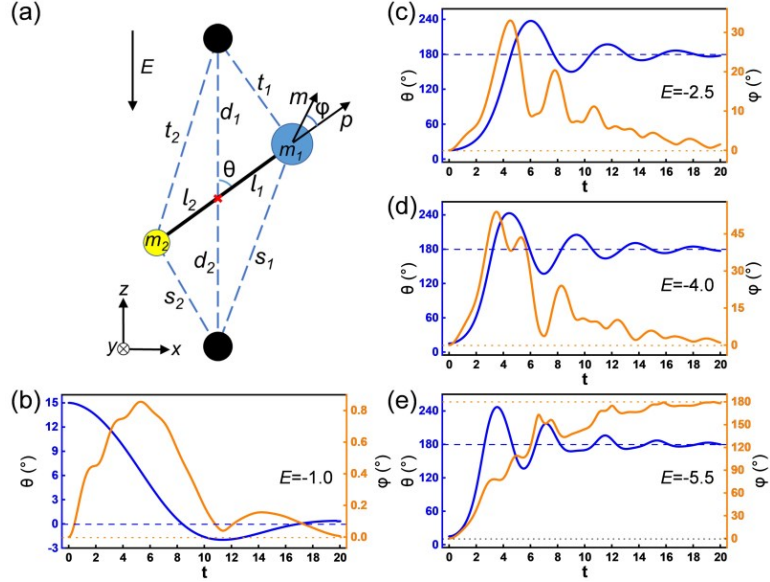

Figure 2. (a) The simplified single molecule multiferroic model and (b)-(e) results of the spin-lattice dynamic simulation. The values of the external electric field in the simulations are set to be (b) $E = -1.0$, (c) $E = -2.5$, (d) $E = -4.0$, and (e) $E = -5.5$. $\theta$ ($\varphi$) is the angle between the electric dipole and the z-axis (magnetic moment). The parameter values of single ion anisotropy are all set to $D = 1.5$ in the simulations of Figure 2b-e.

To characterize the behavior of electric and magnetic dipole moments of the system under external electric fields, as shown in Figures 2b-e, we plot the related angle $\theta$ and $\varphi$ as a function of the simulated time. In these four simulations, electric fields of different magnitudes are applied to the system, but with the same single-ion anisotropy parameter. It is clear that, compared to the results in Figures 2c-e, the small electric field in Figure 2b cannot lead to the reversal of the electric dipole of the diatomic molecule, which implies that a strong enough external electric field is required for the diatomic molecule to overcome the torque contributed by the surrounding atoms and reverse its electric dipole moment. However, as shown in Fig. 2e, an excessively strong electric field is also undesirable, as it would cause the molecule to rotate too fast, making the magnetic moment unable to respond in time. Fortunately, Figures 2c and d show the desirable results, that is when $\theta$ stabilizes to 180°, $\varphi$ stabilizes to 0°, which indicates that the diatomic molecule can not only reverse its electric dipole moment but also generate a well-timed magnetic response to reverse the magnetic moment. Thus, we can conclude that a suitable electric field is required for the appearance of multiferroicity in our proposed single-molecule multiferroic system. Furthermore, the untimely response of the magnetic moment to the electric field in Fig. 2e indicates that the strength of the single ion anisotropy of the diatomic molecule also plays a decisive role in creating multiferroics.

In order to further reveal the role of the single ion anisotropy in our proposed single molecule multiferroics, different values of parameter $D$ are also considered in our simulations. As shown in Figure S1a-d of the SM, we plot the $z$-components of the electric dipole moment and magnetic moment (*i.e.*, $p_z$ and $m_z$, respectively) as a function of the external electric field for different values of the $D$ parameter. Note that some coarse graining treatments are performed on the values of $p_z$ and $m_z$ after extracting their precise values at time t = 30 for each spin-lattice dynamic simulation (see details in Sec. I of SM). In brief, the values +1, -1, and 0 of $p_z$ and $m_z$ in Figure S1 represent electric dipole and magnetic moment aligned parallel, antiparallel, and perpendicular to the $z$-axis, respectively. From Figure S1a-d, we can see that a weak external electric field ($-1.2 < E < 0$) cannot eventually reverse the diatomic molecule, which we have discussed above. Similarly, excessively strong electric fields will finally lead to unpredictable directions of magnetic moment, since the magnetic moment cannot respond to the electric field in time. However, with the increasing strength of the single-ion anisotropy, the desirable interval of the suitable electric field for electric dipole and magnetic moment reversal becomes larger, *i.e.*, -1.2~-1.8, -1.2~-2.4, -1.2~-4.4, and -1.2~-7 for $D$ = 0.5, 1, 1.5, and 2, respectively. Such behavior is due to the fact that a larger $D$ enables the system to confine the magnetic moment with a shorter response time, and thus tolerate a stronger external electric field to reverse the magnetic moment. Then, as shown in Figure S1e-f, we plot $p_z$ and $m_z$ as a function of single ion anisotropy strength for different electric fields. The results also reveal that the stronger the electric field, the larger single-ion anisotropy will be needed to make the magnetic moment reverse with the molecule. Therefore, Figure S1 clearly indicates that the single ion anisotropy does act as a constraint on the direction of the magnetic moment, as we expected.

We note that the electric fields we applied in the abovementioned simulations are static electric fields. One may also ask if the multiferroic character can be achieved under pulse electric fields since (ultrafast) pulse electric fields have shown excellent control on ferroelectricity[23-25], and magnetic property[26,27]. Thus, as displayed in Figure S2 of the SM, we perform the related simulations to demonstrate the ultrafast reversal of the electric dipole and magnetic moment under a single pulse of electric field. One can easily see that, as similar to the abovementioned static electric field calculation, the weak pulse of electric field cannot reverse the electric dipole and magnetic moment of the diatomic molecule, as shown in Figure S2a. The reversal of the electric dipole and magnetic moment can be achieved when an enough large pulse of electric field is applied, as revealed in Figures S2b and c. However, a too-strong pulse of electric field would cause the molecule to rotate too fast, making the magnetic moment unable to respond in time, as displayed in Figure S2d. The single pulse of electric field we applied is shown in the inset of Figure S2d. The detailed simulated results with the varying magnitude of electric field ($E$) and single ion anisotropy parameter ($D$) are reported in Figure S3 of SM.

According to the discussion above, we can conclude that the simplified model proposed in Fig. 2a can indeed exhibit satisfactory performance under the conditions of a sufficiently strong single-ion anisotropy and the application of an appropriate applied electric field, thereby verifying the rationality of our proposed single-molecule multiferroics.

*Realistic Example.* – Now, let us demonstrate the concept of single molecule multiferroics with a realistic example. According to the above discussions, we know that a strong single-ion anisotropy is required for the system to achieve multiferroicity. Therefore, in this part, we first perform DFT calculations on molecules to search for some single molecules with both intrinsic electric dipole moment and strong single-ion anisotropy. We note that although many single molecules with intrinsic electric dipoles and magnetic moments are reported experimentally, most of them are large in size and contain lots of atoms, such as $[Ni(Me_6tren)Cl]^{-}$[28], $[(tpa^{Ph})Fe]^{-}$[29,30], and $[LnDOTA(H_2O)]^{-}$ (Ln=rare earth elements)[31]. Considering the computational cost and the fact that the octahedral and square pyramidal structures are common structural components in various systems, such as $[M(NH_3)_{6-m}X_m]^{n+}$ (M= transition metal elements, X=halogen elements)[32-34] and $[OsNR]^{n-}$ (R=NH_3, C≡CH, Cl)[35], they are adopted as small single-molecule structures with further substituting the elements or molecules. The structures of the single molecules we adopted are shown in Figure 3, and are $M(NH_3)_4N$, $M(NH_3)_5N$, and $MCl_5N$ (M=transition metal elements), respectively. It is evident that there is an intrinsic electric dipole moment in these structures. When performing DFT calculations, the single molecules are placed individually in a unit cell that is large enough to exclude coupling between molecules due to the periodic boundary condition.

Furthermore, the DFT calculations are performed for these structures by only optimizing their atomic positions. The lattice constants of the unit cells are thus fixed in the calculations. Considering the fact that only one molecule exists in the large unit cell, we calculated the magnetic anisotropy energy (MAE) for each structure. In our calculations, structures with different transition elements are tested. As shown in Figure S6 of the SM, we summarize the MAE results for the tested molecules. Among them, the single molecule $Co(NH_3)_4N$ shows a much larger MAE (~1.8 meV/cell, see details in Figure S7 of the SM) compared to other molecules. In Figure S8 of the SM, we show the Lewis structure of $Co(NH_3)_4N$ molecule. Furthermore, we confirm the stability of $Co(NH_3)_4N$ molecule by calculating the formation energy (see section V of the SM). It is worth noting that molecules with even larger MAE can be found[28,36,37]. In other words, while in our study we just prove that our idea of single-molecule multiferroics is feasible, better material systems can continue to be searched in the future. Furthermore, their magnetic moment is parallel to their electric dipole. Therefore, it is adopted as the desired single molecule with intrinsic electric dipole and strong single-ion anisotropy.

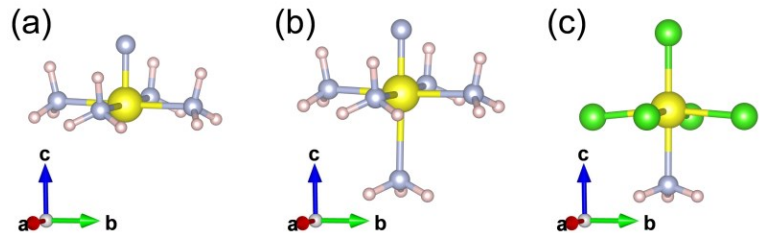

Figure 3. Schematic of the polar single molecules adopted in DFT calculations. The structure in Figure 3a-c has the formula of $M(NH_3)_4N$, $M(NH_3)_5N$, and $MCl_5N$ (M = transition metal elements), respectively. Their electric dipoles lie along the [001] direction of the unit cell.

After obtaining the desired single molecule, the question now becomes how to make it a realistic single-molecule multiferroic material. Here, a strategy is to "encapsulate" the obtained molecules into a single-walled carbon nanotube (SWCNT), as shown in Figure 4. We note that some other materials, such as metal-organic frameworks and fullerenes with large enough internal voids, may also be adopted as hosts, *i.e.*, the "chemical environment" in Figure 1. The SWCNT we adopted has a chiral index of (16, 0) to ensure the semiconducting character. The diameter of the SWCNT is around 12.61 Å, which is larger than that of the molecule (~5.52 Å), therefore, the molecule can easily rotate in the nanotube under external electric fields. The tube is periodical along the *c*-axis with the lattice constant being 12.85 Å. Such a lattice constant is large enough to exclude the interactions between the adjacent molecules, while also ensuring a high-packing density of the single molecules. We name this system as Co(NH$_3$)$_4$N@SWCNT. Furthermore, based on the reported triangular arrangement of the SWCNTs with the tube-tube distance of 3.15 Å[38,39] and assuming that the Co(NH$_3$)$_4$N@SWCNTs are arranged parallel and perpendicular to the substrate, the storage density of the system can be as high as 34.10 TB/in$^2$, which is 1-2 orders of magnitude higher than the reported storage densities using typical storage schemes, such as hard disk drives and magnetic tapes[40-42]. The thickness of the system can be as thin as 12.85 Å, *i.e.*, only one molecule per SWCNT. The electric dipole of the molecule in Fig. 4 directs along the *c*-axis. In order to verify the multiferroicity in such a realistic system, three configurations are conducted with the electric dipole of the embedded molecule directed along the *c*, *b*, and -*c*-axis, respectively. Then, the atomic positions, as well as the lattice constant along the *c*-axis, are fully relaxed through DFT calculations, while the in-plane lattice constants along the *a* and *b*-axes are fixed during the optimization. The structure optimization is performed for all three configurations. The results show that the configuration with electric dipole directed along the +*c* and -*c*-axis are symmetrically identical and are about 0.13 eV/cell lower in energy than the configuration with the electric dipole directed along the *b*-axis. Finally, the magnetic anisotropy energies for the three configurations are calculated to be around 0.8 meV/cell, which indicates that for these three configurations, their magnetic moments remain parallel to the electric dipole moment.

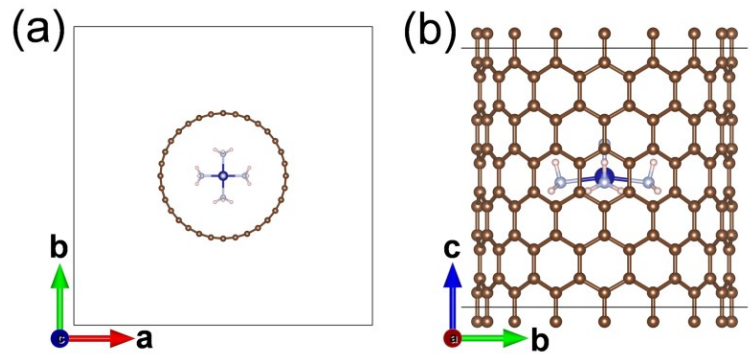

Figure 4. Schematic of the realistic single molecule multiferroic system Co(NH$_3$)$_4$N@SWCNT. The single molecule Co(NH$_3$)$_4$N with intrinsic electric dipole and strong single-ion anisotropy is encapsulated into a single-walled carbon nanotube (SWCNT).

The energy barrier between the two FE states determines the ability of ferroelectricity to switch. As a result, in order to investigate the possible switching process between the two FE states of the system, climbing nudged elastic band (cNEB) calculations[43,44] are performed. As shown in Figure 5, the energy barrier is around 0.13 eV/cell, which is similar to that of conventional ferroelectrics[45-47], and thus can be overcome under an external electric field. We note that the intermediate structures between the two FE states are initially constructed by gradually rotating the molecule about the *a*-axis, thus the electric dipole would gradually rotate from the +*c*-axis to the +*b*-axis and then to the -*c*-axis.

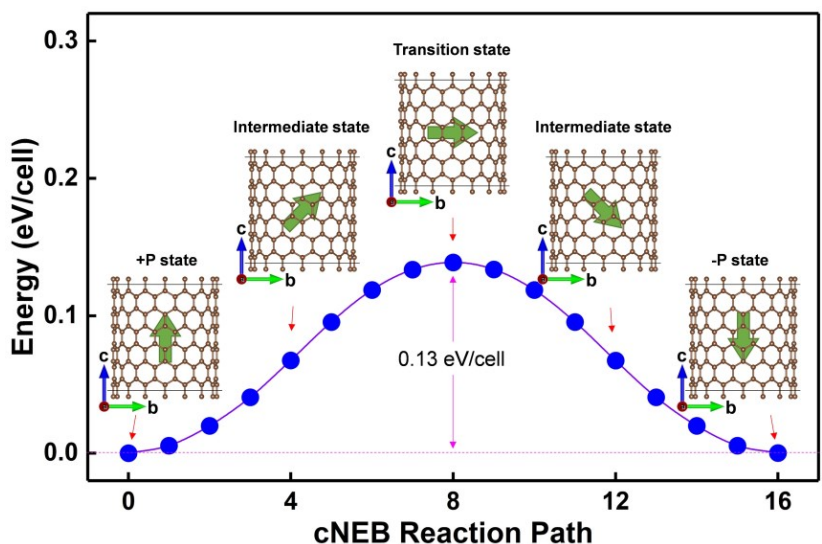

Figure 5. Energy barrier of transition between the FE states with opposite electric dipoles of the realistic single molecule multiferroic system Co(NH$_3$)$_4$N@SWCNT. The green arrows in SWCNT represent the polar molecule Co(NH$_3$)$_4$N with the direction indicating the polarization direction of the molecule.

In order to further verify the correctness of the Hamiltonian model we proposed and the effectiveness of the realistic system we adopted, we extracted the relevant calculation results of the above realistic system, such as the energy barrier (~0.13 eV/cell) of transition between the FE states and the MAE (~0.8 meV/cell) of the Co(NH$_3$)$_4$N@SWCNT system, and then use them as the input parameters of the Hamiltonian model and perform simulations. As shown in Table II, we list detailed information on the realistic parameters adopted in the simulations. The simulated results are shown in Figure 6. It is easy to see that, by using the realistic parameters, we can also achieve the reversal of the electric dipole and the magnetic moment of the molecule when suitable electric fields are applied to the system, which indicates that the realistic system we adopted is indeed capable to realize the idea of single-molecule magnetics. We note that the simulation using a pulse of electric field and realistic parameters are also performed (see Sec. III of the SM), and that the results show that the reversal of the electric dipole and the magnetic moment of the system can also be achieved. Furthermore, the simulations considering the temperature effect on the realistic system are also performed (see details in Sec. IV of the SM), and they reveal that the single-molecule multiferroic behavior of the realistic system can exist under finite temperature.

TABLE II. The parameters of the realistic model for Co(NH$_3$)$_4$N@SWCNT adopted in the spin-lattice dynamic simulations. We note that $\theta_0$ denotes $\theta_{t=0}$ and $\delta t$ represents the time step of the dynamic simulations.

| Parameter | $l_1$ (Å) | $l_2$ (Å) | $d_1$ (Å) | $d_2$ (Å) | $m_1$ (u) | $m_2$ (u) | $A$ (eV·Å$^6$) |
|---|---|---|---|---|---|---|---|
| Value | 2.0 | 2.0 | 6.0 | 6.0 | 28.0 | 28.0 | 300.0 |
| Parameter | $p$ (e·Å) | $m$ ($\mu_B$) | $D$ (meV) | $\alpha$ | $\theta_0$ (°) | $\delta t$ (fs) | $C$ (eV·fs) |
| Value | 2.0 | 2.0 | 0.8 | 0.01 | 15.0 | 0.1 | 4.0 |

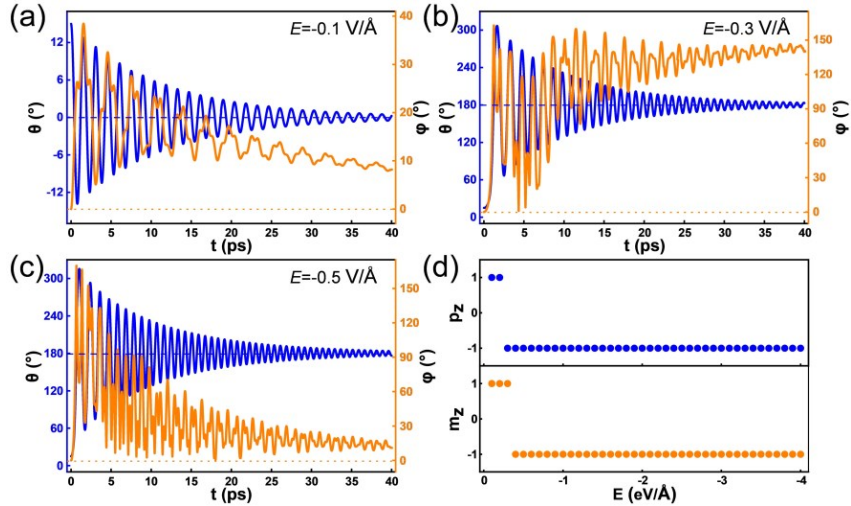

Figure 6. Results of the spin-lattice dynamics simulation at $T = 0$ K with real parameters. $\theta$ ($\varphi$) is the angle between the electric dipole and the $z$-axis (magnetic moment). The values of the adjustable parameters in each case are set to (a) $E = -0.1$ V/Å, (b) $E = -0.3$ V/Å, (c) $E = -0.5$ V/Å, (d) $E = -0.1 \sim -4$ V/Å.

**Conclusion**

To summarize, we achieve a rational design of single-molecule multiferroics which should lead to the ultimate memory. To this end, a simplified model, as well as the total Hamiltonian, are built and reveal that the single-molecule multiferroicity can arise in systems where the molecules have intrinsic electric dipoles and large single-ion anisotropy. The magnetic moment of the system can be switched with the reversal of electric polarization under a suitable external electric field and strong single-ion anisotropy. Taking advantage of the DFT calculations, we demonstrate that the embedding of a multiferroic single molecule with large single-ion anisotropy into a single-walled carbon nanotube represents a plausible and realistic example. The energy barrier between the two ferroelectric states is estimated to be similar to that of conventional ferroelectrics. This work therefore not only sheds light on the understanding and discovery of new single-molecule multiferroics but also provides a new guideline to design multiferroics for potential usage in high-data-density and low-energy-consuming storage.

**Supporting Information**

Spin-lattice dynamics simulations under static electric fields; Spin-lattice dynamics simulations under a pulse of electric fields; Spin-lattice dynamics simulations with real parameters and under a pulse of electric fields; Spin-lattice dynamics simulations with real parameters and at finite temperatures; DFT-calculated MAE results for the tested single molecules.

**Author Information**

**Corresponding Author**

*E-mail: hxiang@fudan.edu.cn

**Author Contributions**

†Y.Y. and L.H. contributed equally to this work.

**Notes**

The authors declare no competing financial interest.

**Acknowledgments**

The work at Fudan is supported by the National Key R&D Program of China (No. 2022YFA1402901), NSFC (grants No. 11825403, 11991061, 12188101), and the Guangdong Major Project of the Basic and Applied Basic Research (Future functional materials under extreme conditions--2021B0301030005). L.B. acknowledges the support by the Vannevar Bush Faculty Fellowship (VBFF) Grant No. N00014-20-1-2834 from the Department of Defense and Award No. DMR-1906383 from the National Science Foundation AMASE-i Program (MonArk NSF Quantum Foundry).